\documentclass[prb,amssymb,twocolumn,showpacs,superscriptaddress,floatfix]{revtex4}

\usepackage{graphicx}
\usepackage{bm}
\usepackage{mathptmx}

\begin{document}

\title{Fractal dimension of domain walls in the Edwards-Anderson spin glass
model}

\author{S. Risau-Gusman}
\affiliation{Centro At{\'{o}}mico Bariloche, R8402AGP San Carlos
de Bariloche, R\'{\i}o Negro, Argentina}
\author{F. Rom\'a}
\affiliation{Centro At{\'{o}}mico Bariloche, R8402AGP San Carlos
de Bariloche, R\'{\i}o Negro, Argentina} \affiliation{Departamento
de F\'{\i}sica, Universidad Nacional de San Luis, Chacabuco 917,
D5700BWS San Luis, Argentina}
\date{\today}

\begin{abstract}
We study directly the length of the domain walls (DW) obtained by
comparing the ground states of the Edwards-Anderson spin glass
model subject to periodic and antiperiodic boundary conditions.
For the bimodal and Gaussian bond distributions, we have isolated
the DW and have calculated directly its fractal dimension $d_f$.
Our results show that, even though in three dimensions $d_f$ is
the same for both distributions of bonds, this is clearly not the
case for two-dimensional (2D) systems. In addition, contrary to
what happens in the case of the 2D Edwards-Anderson spin glass
with Gaussian distribution of bonds, we find no evidence that the
DW for the bimodal distribution of bonds can be described as a
Schramm-Loewner evolution processes.

\end{abstract}


\date{\today}

\maketitle

Whereas the properties of long range Ising spin glasses are now
well understood, after more than 20 years of work, the same cannot
be said about short range spin glasses. This is true even for very
simple models: the nature of the ordering of the low temperature
phase of the two-dimensional (2D) Edwards-Anderson (EA) spin glass
model \cite{EA} is still being debated. Even though the fact that
at $T=0$ EA models with Gaussian (EAG) and bimodal (EAB) bond
distributions belong to two different universality classes seems
well established, \cite{Amoruso2003} new studies
\cite{Hartmann2007} of low energy excitations (fractal droplets)
show that there is still room for discussion.

It has recently been suggested \cite{Amoruso2006,Bernard2007} that
domain walls (DWs) can be described as Schramm-Loewner evolution
(SLE) processes. These processes are Brownian walks of diffusion
constant $\kappa$ and fractal dimension $d_f=1+ \kappa/8$.
Furthermore, using conformal field theory the stiffness exponent
$\theta$ (which characterizes the scaling of the DW energy) can be
related to the fractal dimension $d_f$ via
\begin{equation}
d_f^{SLE} = 1+ \frac{3}{4(3+\theta)}. \label{SLE}
\end{equation}
This seems to be true for the 2D EAG, as for $\theta = -0.287(4)$
\cite{Hartmann2002} Eq.~(\ref{SLE}) gives ${d_f}^{SLE}
=1.2764(4)$, which is compatible with the best numerical estimate
$d_f=1.274(2)$. \cite{Melchert2007} It is not clear, however,
whether such a relation should hold for the 2D EAB, because of the
high degeneracy of its ground state (GS). If it did, using the
fact that the stiffness exponent seems to vanish, \cite{Hartmann5}
Eq.~(\ref{SLE}) would yield ${d_f}^{SLE} =1.25$.

In the EAB model the degeneracy of the GS precludes a clear-cut
definition of the fractal dimension of the DW. For this reason,
most of the estimates of $d_f$ are based on a scaling argument of
Fisher and Huse, \cite{Fisher1} which states that the entropy of
droplets of size $L$ should scale as $S_{DW} \sim L^{d_f/2}$. It
must be stressed that this was originally proposed for systems
with only one GS. The estimates obtained using this scaling range
from $d_f \approx 1.0$ to $d_f=1.30(3)$.
\cite{Saul1993,Aromsawa2007} Even though very recently more direct
measurements have been attempted, \cite{Roma2007,Weigel2007} in
those works the sampling of the DWs was not controlled.
\cite{Melchert2007} In Ref.~\onlinecite{Melchert2007} this problem
is avoided and bounds are provided for the true $d_f$:
$1.095(2)<d_f<1.395(3)$.

In this paper we present the results of an extensive numerical
study of the fractal dimension of DWs, using a direct measure of
their length. We have studied both 2D and 3D systems with Gaussian
and bimodal distributions of bonds, but we have concentrated on
the EAB model. For small systems we have calculated the exact
average DW length. For larger systems an estimate of this quantity
has been calculated by choosing only one pair of GSs. Even though
the sampling we use is clearly not uniform, the values we obtain
with this estimate coincide with the exact ones, for small
systems. Our results show that, whereas in 3D $d_f$ is the same
for EAB and EAG, in the 2D case the corresponding values are
clearly different and, in particular, for the EAB, different from
the value that would be obtained from Eq.~(\ref{SLE}) (assuming
that $\theta=0$). In addition, we have performed one test to see
whether DWs in 2D can be described as SLE processes. Even though
for the EAG the result of this test is positive, for the EAB the
outcome of the test is clearly negative.

We start by considering the Hamiltonian of the EA model for spin
glasses \cite{EA} on square and cubic lattices, $H = \sum_{( i,j
)} J_{ij} \sigma_{i} \sigma_{j}$, where $\sigma_i = \pm 1$ is the
spin variable and $( i,j )$ indicates a sum over nearest
neighbors. The coupling constants are independent random variables
chosen either from a bimodal ($\pm J$) or Gaussian bond
distributions, both with zero mean and variance one.
\begin{figure}[hb]
\includegraphics[width=\linewidth,clip=true]{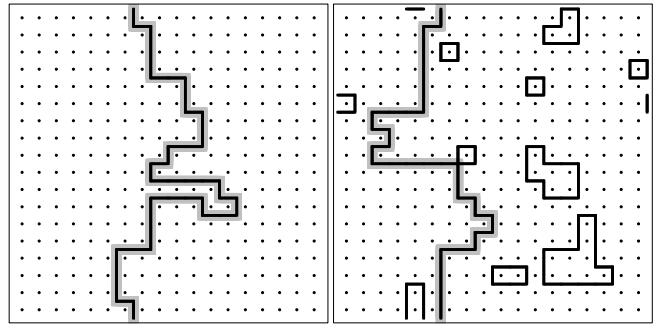}
\caption{\label{figure1} DWs for two 2D samples with different
bonds distributions and periodic BCs in both directions. Points
are spins. The lines cross the bonds (not shown) that have
`changed state' when antiperiodic BCs are introduced in the
horizontal direction. The thick line is the DW. Left: Gaussian.
Right: Bimodal. }
\end{figure}
In 3D we have used periodic boundary conditions (BCs) in all
directions, and in 2D we have studied two cases: periodic-periodic
BCs and periodic-free BCs.

For the EAG the definition of the DW is straightforward: given a
system we `perturb' it by changing the sign of all the bonds in a
column in 2D or plane in 3D (i.e. we introduce antiperiodic BCs in
one direction) and compare the GS of the new system with that of
the unperturbed one. The DW is simply the set of bonds which have
`changed state', i.e. those for which $J_{ij} \sigma_{i}
\sigma_{j}$ has changed its sign. It can be shown that in the dual
lattice, the dual of the bonds that have changed can only form a
path in 2D (or surface in 3D) \cite{notadual} of length (or area)
$l$, that goes from one border of the lattice to the opposite one
(see Fig.~\ref{figure1}). $d_f$ is obtained from the relation
$l_{DW} = \langle l \rangle \sim L^{d_f}$, where $l_{DW}$ is the
sample average ($\langle \cdots \rangle$) of $l$.

\begin{table}
\caption{\label{t1} Parameters used in the MC runs for 2D systems
(See text for details). }
\begin{ruledtabular}
\begin{tabular}{ccccccc}
&EAB&& \ \ \ &&EAG&\\
$L$&$t$&$N_S$& \ \ \ &$L$&$t$&$N_S$ \\
\hline
$3$ & $2\times10^3$ & $10^4$ & \ \ \ & $3$ & $2\times10^3$ & $10^4$  \\
$4$ & $2\times10^3$ & $10^4$ & \ \ \ & $4$ & $2\times10^3$ & $10^4$  \\
$5$ & $2\times10^3$ & $10^4$ & \ \ \ & $5$ & $2\times10^3$ & $10^4$  \\
$6$ & $2\times10^3$ & $10^4$ & \ \ \ & $6$ & $2\times10^3$ & $10^4$  \\
$7$ & $2\times10^3$ & $10^4$ & \ \ \ & $7$ & $2\times10^3$ & $10^4$  \\
$8$ & $2\times10^3$ & $10^4$ & \ \ \ & $8$ & $4\times10^3$ & $10^4$  \\
$9$ & $2\times10^3$ & $10^4$ & \ \ \ & $9$ & $10^4$ & $10^4$  \\
$10$ & $2\times10^3$ & $10^4$ & \ \ \ & $10$ & $2\times10^4$ & $10^4$  \\
$12$ & $4\times10^3$ & $10^4$ & \ \ \ & $12$ & $10^5$ & $5\times10^3$  \\
$14$ & $10^4$ & $10^4$ & \ \ \ & $14$ & $4\times10^5$ & $3\times10^3$  \\
$16$ & $4\times10^4$ & $10^4$ & \ \ \ & $16$ & $10^6$ & $2\times10^3$  \\
$18$ & $8\times10^4$ & $6\times10^3$ & \ \ \ & $18$ & $3\times10^6$ & $10^3$  \\
$20$ & $2\times10^5$ & $3\times10^3$ & \ \ \ &  &  &   \\
$22$ & $4\times10^5$ & $10^3$ & \ \ \ &  &  &   \\
\end{tabular}
\end{ruledtabular}
\end{table}

\begin{table}
\caption{\label{t2} Parameters used in the MC runs for 3D systems
(See text for details). }
\begin{ruledtabular}
\begin{tabular}{ccccccc}
&EAB&& \ \ \ &&EAG&\\
$L$&$t$&$N_S$& \ \ \ &$L$&$t$&$N_S$ \\
\hline
$2$ & $2\times10^3$ & $10^4$ & \ \ \ & $2$ & $2\times10^3$ & $10^4$  \\
$3$ & $2\times10^3$ & $10^4$ & \ \ \ & $3$ & $2\times10^3$ & $5\times10^3$  \\
$4$ & $2\times10^3$ & $10^4$ & \ \ \ & $4$ & $2\times10^3$ & $5\times10^3$  \\
$5$ & $2\times10^3$ & $6\times10^3$ & \ \ \ & $5$ & $6\times10^3$ & $3\times10^3$  \\
$6$ & $4\times10^4$ & $3\times10^3$ & \ \ \ & $6$ & $4\times10^4$ & $2\times10^3$  \\
$7$ & $2\times10^5$ & $10^3$ & \ \ \ & $7$ & $2\times10^5$ & $10^3$  \\
$8$ & $6\times10^5$ & $10^3$ & \ \ \ & $8$ & $10^6$ & $10^3$  \\
$9$ & $2\times10^6$ & $10^3$ & \ \ \ & $9$ & $4\times10^6$ & $7\times10^2$  \\
$10$ & $10^7$ & $5\times10^2$ & \ \ \ & $10$ & $2\times10^7$ & $5\times10^2$  \\
\end{tabular}
\end{ruledtabular}
\end{table}

\begin{figure}
\includegraphics[width=7.5cm,clip=true]{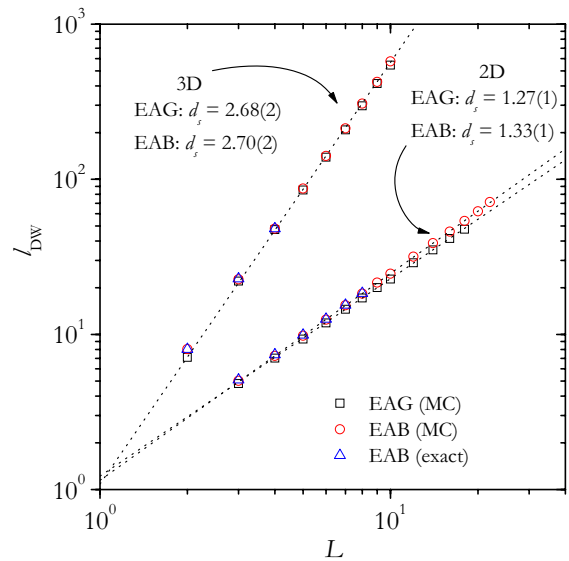}
\caption{\label{figure2} (color online) DW length for both
Gaussian and bimodal distribution of bonds in 2D and 3D for
systems with periodic boundary conditions in all directions. The
exact values were obtained with a branch-and-bound algorithm.
Error bars are smaller than the symbols. }
\end{figure}

For the EAB one has to be a bit more careful because the GS is
degenerated. If one performs the same procedure as above, but now
with a random pair of GSs, one sees that {\it many} sets of bonds
have changed state (see Fig.~\ref{figure1}). Only one of these
sets forms a structure that crosses the sample, as in the EAG. The
other sets form loops of zero energy: they enclose clusters of
spins that can be flipped without changing the energy of the
sample. As these clusters are present both in the perturbed and
the unperturbed systems, it is reasonable to define the DW of a
given pair of GS, as only the structure that crosses the sample.
In this case the characteristic length of the DW of a sample, $l$,
is defined as the average over all pair of GSs of the length of
DWs, as defined above.  In practice, to extract the DW in the EAB
one has to look for the percolating structure in the dual lattice,
both for 2D and 3D. For this we have used the algorithm of
Hoshen-Kopelman. \cite{Hoshen1976} We have been careful to
eliminate the loops (or closed surfaces) that stick to the DW (see
Fig.~\ref{figure1} for an example).

For small EAB samples we have obtained the exact value of the DW
length by averaging over all pairs of GSs for each sample. To
obtain all the GSs for systems with fully periodic BCs we have
used a branch-and-bound algorithm. \cite{Hartwig1984} We have
studied sizes up to $L=8$ in 2D and $L=4$ in 3D ($10^4$ samples
for each size). Results are shown in Fig.~\ref{figure2}. This
approach is not practical for larger sizes because of the fact
that the number of GSs grows exponentially with $L$. \cite{Landry}
Therefore, to estimate the average DW length for larger samples,
we have resorted to a Monte Carlo (MC) algorithm to obtain one
pair of random GSs for each sample (see Fig.~\ref{figure2}).

The MC algorithm we have used is a variant of parallel tempering
\cite{Hukushima1996} but suitably modified to find a ground state
as quickly as possible. As in Ref.~\onlinecite{Hukushima1996}, we
use a compound system or ensemble, that consists of $m$
noninteracting replicas of the system, each one associated to a
different temperature in the interval $[T_{min},T_{max}]$, where
the distance between consecutive temperatures is a constant. In
general the heart of a parallel tempering algorithm consists of
two routines that are performed alternately. One of them consists
of a standard MC algorithm applied to each replica: in each
elementary step, the update of a random selected spin of the
ensemble is attempted with probability given by the Metropolis
rule \cite{Metropolis}. In the other routine, an exchange of two
configurations between two replicas at consecutive temperatures is
attempted with the probability defined in
Ref.~\onlinecite{Hukushima1996}. In general, the unit time or MC
step (MCS) of a parallel tempering algorithm consists of a fixed
number of elementary steps of standard MC, followed by another
fixed number of trials of replica exchange. To equilibrate the
system a MCS defined as $m\times N$ elementary steps of standard
MC and only one replica exchange, is usually chosen. As we are
only interested in reaching quickly a configuration of GS, we have
used a different MCS: it consists of $m\times N$ cycles, where a
cycle is defined as only one elementary step of standard MC plus
one replica exchange. After $t$ MCS, the algorithm stops and the
configuration with minimum energy (among all configurations
visited in all replicas in the simulation process) is stored. We
have found that this algorithm is very efficient for reaching the
GS: we have checked that, for the sizes studied with the
parameters indicated in Tables I and II, our heuristic outputs a
true GS with a probability larger than $0.99$. In particular for
2D systems, we have verified this by calculating the ground states
energies with an exact branch-and-cut algorithm
\cite{DeSimone,server} and comparing them with the energies of the
configurations we obtained (a detailed analysis of this algorithm
will be published elsewhere \cite{Roma2008}).

In our simulations, for each lattice size $L$ of EAB and EAG in
both 2D and 3D systems we have used $m=20$ replicas with
temperatures between $T_{max}=1.6$ to $T_{min}=0.1$. The number of
MCS $t$ and the number of samples $N_S$ analyzed for each sample
size, is given in Tables I and II.

\begin{figure}
\includegraphics[width=7.5cm,clip=true]{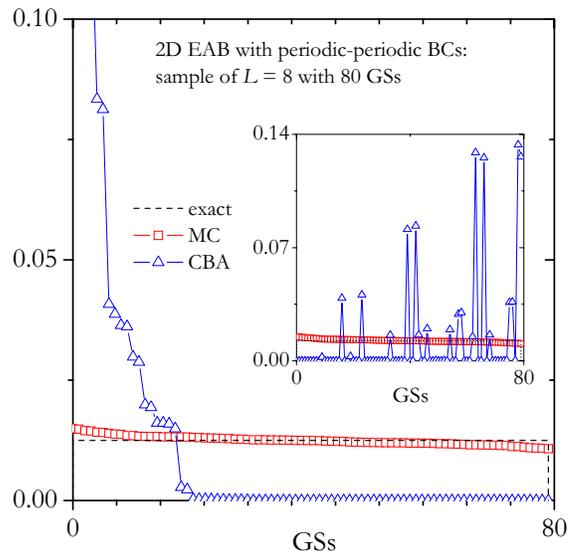}
\caption{\label{figure3} (color online) Histogram of the fraction
of runs for which each GS has been obtained, for two different
stochastic algorithms (see text). The states have been ordered by
decreasing frequency of sampling, for each algorithm, i.e. the
orderings depend on the algorithm. In the inset the ordering is
the same for both algorithms. }
\end{figure}

Interestingly, considering only one pair of GSs for each EAB
sample leads to an estimate of the average DW length that is
within only $1\%$ of its exact value (for small systems). This is
good evidence that the MC algorithm is sampling the GS space
almost uniformly. As our algorithm does not reach equilibrium we
have checked explicitly that the sampling is indeed almost
uniform. In Fig.~\ref{figure3} we show the distribution of GSs
reached by our algorithm for a typical 2D sample with 80 GSs. It
is instructive to compare our method with the sampling that
results from the algorithm (hereafter called CBA) used by Cieplak
and Banavar. \cite{Cieplak} In this technique an infinitesimal
noise is added to the couplings, and the GS of the system is
obtained. Then, the same realization of the noise is added to the
system with antiperiodic BCs and a new GS is obtained. The noise
in the couplings breaks the degeneracy, and the fact that the same
realization of the noise is used ensures that the resulting DW has
no loops. Up to $L=8$ we have also used the CBA to analyze the
same samples as with the branch-and-bound algorithm. As
Fig.~\ref{figure3} shows, the sampling given by this technique is
far from uniform. In fact, there is a large fraction of GSs that
are not reached by it. In spite of this, the estimate of the
average DW length given by the CBA is within $3\%$ of its exact
value (it must be stressed that the bias is systematic: the
estimates  are consistently smaller than the exact values). This
suggests that the sampling method might not be crucial to obtain
good estimates of $d_f$.

For 3D EAG systems (see Fig.~\ref{figure2}) the value we obtain is
$d_f=2.68(2)$. This value is compatible with the ones found by
other authors: $d_f=2.68(2)$ and $d_f \approx 2.7$.
\cite{PalYou99,Aspelmeier} However, the fractal dimension obtained
when comparing GSs obtaining by studying the response of the GS to
a coupling-dependent bulk perturbation is a bit smaller:
$d_f=2.57(2)$. \cite{Palassini2003} This can perhaps be explained
by arguing that, given the fact that droplets of all sizes appear
when there is a bulk perturbation, the fractal dimension of all
the droplets do not have to coincide. Recently it has been found
that in 2D droplets with different sizes do seem to have different
values of $d_f$. \cite{Hartmann4}

For 3D EAB systems we have compared the average DW lengths
obtained using our MC algorithm and the CBA algorithm, up to
$L=7$. As in the 2D case, the difference between both values is
almost constant and smaller than $1\%$. Using our MC algorithm up
to $L=10$ we obtain $d_f=2.70(2)$. A different value is obtained
in Ref.~\onlinecite{Roma2007}, where the DW is restricted to the
{\it rigid lattice}. This is the set of bonds which do not change
its state in all the GSs (i.e. the bonds are either frustrated or
unfrustrated in all GSs). The value reported there for 3D
($d_f=2.59(2)$) is a bit smaller than the one reported in this
article. This is due to finite size effects, given the fact that
the size of the rigid lattice varies greatly from sample to
sample.

For 2D systems we obtain $d_f=1.33(1)$ for the EAB and
$d_f=1.27(1)$ for the EAG (see Fig.~\ref{figure2}). Notice that
these two values are clearly different. Given the small systems
considered, this discrepancy could be attributed to finite size
effects (which are usually large in 2D systems). As much larger
sizes can be analyzed in lattices with one free BC, we have
studied the length of the DW in those systems. It is well known
\cite{Hartmannlibro} that in this case the problem of finding the
GS can be mapped to a minimum-weighted perfect matching problem,
for which very efficient algorithms exist. We have used one
implementation of a Blossom algorithm, \cite{Blossom} which has
allowed us to obtain the GSs up to $L=300$ for EAB and up to
$L=100$ for EAG ($10^3-10^4$ samples for each size).

\begin{figure}
\includegraphics[width=7.5cm,clip=true]{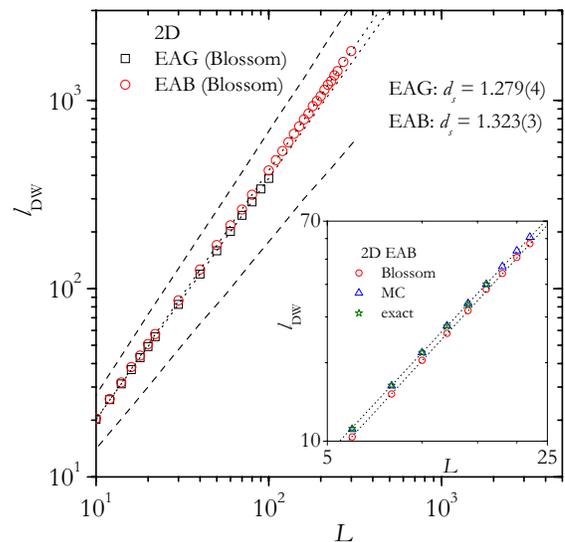}
\caption{\label{figure4} (color online) DW lengths for 2D systems
with periodic-free BCs. The dashed line represent the bounds found
in Ref.~\onlinecite{Melchert2007}. The inset shows both the exact
average and the Blossom estimate of the DW length, as well as a
Monte Carlo estimate, for samples of small size.  Error bars are
smaller than the symbols.}
\end{figure}

As in the case of systems with fully periodic BCs, for the EAB we
have calculated the exact average length of the DW for small
systems (up to $L=16$). For this we have used an algorithm to
count all the minimum-weighted perfect matchings of each sample.
\cite{Landry2002} We have also applied the MC algorithm, which has
allowed us to reach sizes up to $L=22$. Again, the MC estimate is
within $1 \%$ of the exact value. To reach even larger sizes we
have used an algorithm based in the Blossom routine to choose a
pair of random GSs. This was done by choosing a GS among all the
GSs compatible with a given random optimal matching. The results
are shown in Fig.~\ref{figure4}. In the inset the three approaches
are compared, for small systems. It can be seen that the estimate
given by the Blossom algorithm is clearly smaller than the exact
value. This is evidence that not all GSs are reached with the same
probability.

To understand the origin of this bias it is necessary to notice
that each optimal matching corresponds to many GSs. We call this
the {\it degeneracy} of the matching. But these degeneracies vary
widely from matching to matching. Therefore, a random sampling of
the optimal matchings tends to favor GSs present in matchings of
small degeneracies. For some reason, these particular GSs of the
periodic and antiperiodic system are more similar than GSs chosen
uniformly at random. This leads naturally to smaller DWs. However,
the bias introduced by this nonuniform sampling is so small (less
than $6 \%$) that it is difficult to make this reasoning more
precise (see the inset in Fig.~\ref{figure4}). In spite of this
bias, the exponent estimated with this procedure is very close to
the value given by the exact and MC algorithms, up to $L=22$.
Thus, as in the case of our MC algorithm and the CBA, the only
effect of the bias is to shift the points, but without changing
the slope of the fit. This shows again that the sampling method
does not seem to be crucial to determine the fractal dimension.

Studying systems of sizes up to $L=300$ the value that we estimate
for the fractal dimension of the 2D EAB model is $d_f=1.323(3)$.
This value is between the bounds given in
Ref.~\onlinecite{Melchert2007} and is compatible both with the
value obtained for the case of fully periodic BCs, and with the
most recent estimate calculated using the entropy ansatz:
$d_f=1.30(3)$. \cite{Aromsawa2007} Even though a value of
$d_f=1.30(1)$ has been reported previously, \cite{Roma2007} the
small discrepancy with the value reported in this letter is due to
the larger system sizes that are considered here (up to $L=100$ in
Ref.~\onlinecite{Roma2007}). For the sake of comparison, we have
also calculated the fractal dimension for the 2D EAG. We obtain
$d_f=1.279(4)$, which is compatible with the most recent and
accurate value reported: $d_f=1.274(2)$. \cite{Melchert2007}

Notice that the fractal dimensions obtained for the 2D EAB and the
2D EAG are clearly different. To check that this is not an
artifact of the scaling function used to fit the points, $B
L^{d_f}$, we have also tried fits with two other scaling functions
with additional correction terms (see Table III). For EAB we have
obtained $d_f>1.32$ for all the fits with $Q>0.4$ and $Q>0.1$ for
periodic-free and periodic-periodic systems, respectively (a fit
with $Q>0.1$ is considered as a good fit \cite{numerical}). This
shows that, with a high probability, the difference found between
the fractal dimensions for the EAB and the EAG is not an effect of
finite size scaling. This difference is further evidence that the
universality of both models is different for $T=0$.

\begin{table*}
\caption{\label{t3} Fractal dimension obtained using different
scaling corrections.  The fits are restricted to the interval
$[L_{min},L_{max}]$.  $Q$ gives the quality of the fits (see the
text). }
\begin{ruledtabular}
\begin{tabular}{ccccccccc}
    &     & periodic-periodic BCs&      & \ \ \ &     & periodic-free BCs &   &\\
    &$d_f$& $L_{min} - L_{max}$  &$Q$   & \ \ \ &$d_f$&$L_{min} - L_{max}$&$Q$& \\
\hline
$B L^{d_f}$     &$1.33(1)$&$8 - 22$ &$0.14$& \ \ \ & $1.323(3)$ &$20 - 300$ &$0.57$  \\
$A+B L^{d_f}$   &$1.4(1)$ &$10 - 22$&$0.1$ & \ \ \ & $1.34(2) $ &$60 - 300$ &$0.4$  \\
$A L+ B L^{d_f}$&$1.8(2)$ &$10 - 22$&$0.002$& \ \ \ & $1.40(7) $ &$60 - 300$ &$0.4$  \\
\end{tabular}
\end{ruledtabular}
\end{table*}

As mentioned above, Eq.~(\ref{SLE}) (considering $\theta=0$)
predicts $d_f^{SLE}=1.25$. This is the same universality class as
the loop-erased random walks. \cite{Schramm2000} Interestingly,
the fractal dimension we obtain is much closer to $d_f=4/3$, which
is the fractal dimension of the normal self-avoiding walks.
\cite{Flory} This does not necessarily mean that DWs cannot be
described as SLE processes. The walks generated by these processes
have many interesting properties. One way to see whether DWs can
be described as SLE processes is to test whether they have some of
these properties. \cite{Bernard2007}

\begin{figure}
\includegraphics[width=7.5cm,clip=true]{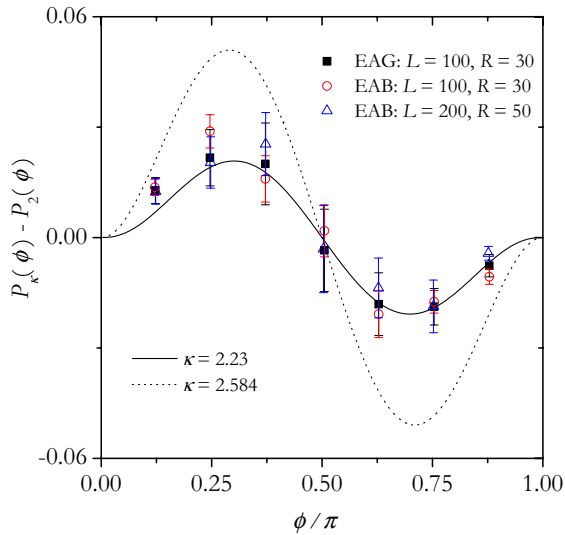}
\caption{\label{figure5} (color online) Comparison between the
prediction of Eq.~(\ref{Pk}) and the values obtained numerically
for 2D EAB and 2D EAG systems. The lines show the difference
between $P_{\kappa} (\phi)$ and $P_2 (\phi)$ for two values of
$\kappa$. The points are averages over 5000 samples for EAB
systems with $L=100$ and $R=30$, and over 2000 samples in the
other cases. }
\end{figure}

We have performed one such test in 2D. If DWs are described by an
SLE process of diffusion constant $\kappa$, the probability that
they pass to the right of a point with polar angle $\phi$
(measured from the starting point of the walk) is
\cite{Schramm2001}:
\begin{equation}
P_{\kappa} (\phi) = \frac{1}{2}-
\frac{{\Gamma}(\frac{4}{\kappa})\cot (\phi)}{\sqrt{\pi}
{\Gamma}(\frac{8-\kappa}{2 \kappa})} {_2F_1}
\biggl[\frac{1}{2},\frac{4}{\kappa};\frac{3}{2};-\cot^2
(\phi)\biggr] \label{Pk}
\end{equation}
where $_2F_1$ is the hypergeometric function. Notice that
$P_{\kappa} (\phi)$ does not depend on the radial coordinate $R$
of the reference point. The result of this test is shown in
Fig.~\ref{figure5}. The points for the EAG are compatible with a
diffusion constant $\kappa = 2.23$, which is consistent with the
value obtained by Bernard {\em et al.}. \cite{Bernard2007} In
turn, this value is compatible with the relation $d_f=1+\kappa /
8$. For the EAB the points obtained are far from the curve (upper
curve in Fig.~\ref{figure5}) that corresponds to
$\kappa=8(d_f-1)=2.584$ (using $d_f=1.323)$. In fact, they are
much closer to $P_{2.23}(\phi)$ (lower curve in
Fig.~\ref{figure5}). Thus, unless finite size effects are very
significant, this result shows that EAB DWs cannot be consistently
described as SLE processes. It must be stressed that this test has
been performed using the DWs calculated with the Blossom algorithm
which means that, for the EAB, the GSs have not been sampled
uniformly. Nevertheless, as the determination of the fractal
dimension of the DWs does not seem to be affected by this bias, it
is not unreasonable to assume that it also does not affect the
outcome of the SLE test. We have only used the Blossom algorithm
because, for this kind of tests, very large system sizes must be
studied to obtain meaningful results.

In conclusion, we have performed an extensive numerical study of
the domain walls for the EAB and the EAG both in 2D and 3D, and
have determined their fractal dimension from a direct measurement
of their length. In 3D we find that both exponents coincide. On
the other hand, for 2D systems we find a clear difference between
the fractal dimensions of the EAB and the EAG. The value obtained
for the EAB model shows that these domain walls are more similar
to self-avoiding walks than to loop-erased random walks, as one
would expect from Eq.~(\ref{SLE}). Even though the methods we have
used do not sample uniformly the ground state space, the fact that
all the obtained values of $d_f$ coincide is good evidence that
this estimate must be very close to its true value: given the very
different nature of the algorithms used it would be very unlikely
that they give raise to the same wrong estimate of the fractal
dimension. \cite{nota} Finally we have tested whether the 2D EAB
domain walls can be described by SLE processes. The outcome of
this test is negative: the probability that a DW passes to the
right of a given point is not consistent with an SLE process with
$\kappa=8(d_f-1)$.

This work has been supported by Universidad Nacional de San Luis
(Argentina) under project 322000, CONICET (Argentina) under
project PIP 6294 and the National Agency of Scientific and
Technological Promotion (Argentina) under project 33328 PICT 2005.


\begin{thebibliography}{99}
\bibitem{EA} S.F. Edwards and P.W. Anderson, J. Phys. F {\bf 5}, \rm 965
(1975).
\bibitem{Amoruso2003} C. Amoruso, E. Marinari, O. C. Martin, and A. Pagnani, Phys. Rev. Lett. {\bf 91}, 087201 (2003).
\bibitem{Hartmann2007} A. K. Hartmann, cond-mat/0704.2748.
\bibitem{Amoruso2006} C. Amoruso, A. K. Hartmann, M. B. Hastings, and M. A. Moore, Phys. Rev. Lett. {\bf 97}, 267202 (2006).
\bibitem{Bernard2007} D. Bernard, P. Le Doussal, and A. A. Middleton, Phys. Rev. B {\bf 76}, 020403(R) (2007).
\bibitem{Hartmann2002} A. K. Hartmann, A. J. Bray, A. C. Carter, M. A. Moore, and A. P. Young, Phys. Rev. B {\bf 66}, 224401 (2002).
\bibitem{Melchert2007} O. Melchert and A. K. Hartmann, Phys. Rev. B {\bf 76}, 174411 (2007).
\bibitem{Hartmann5} A. K. Hartmann and A. P. Young, Phys. Rev. B {\bf 64}, 180404(R) (2001).
\bibitem{Fisher1} D. S. Fisher and D. A. Huse, Phys. Rev. Lett. {\bf 56}, 1601 (1986).
\bibitem{Saul1993} L. Saul and M. Kardar, Phys. Rev. E {\bf 48}, R3221 (1993).
\bibitem{Aromsawa2007} A. Aromsawa and J. Poulter, Phys. Rev. B {\bf 76}, 064427 (2007).
\bibitem{Roma2007} F. Rom\'a, S. Risau-Gusman, A. J. Ramirez-Pastor, F. Nieto, and E. E. Vogel, Phys. Rev. B {\bf 75}, 020402(R) (2007).
\bibitem{Weigel2007} M. Weigel and D. Johnston, Phys. Rev. B {\bf 76}, 054408 (2007).
\bibitem{notadual} Notice that in 2D the bonds in the lattice correspond to bonds in
its dual lattice, but in 3D they correspond to plaquettes.
\bibitem{Hoshen1976} J. Hoshen and R. Kopelman, Phys. Rev. B {\bf 14}, 3438 (1976).
\bibitem{Hartwig1984} A. Hartwig, F. Daske, and S. Kobe, Comput. Phys. Commun. {\bf 32}, 133 (1984).
\bibitem{Landry} J. W. Landry and S. N. Coppersmith, Phys. Rev. B {\bf 65}, 134404 (2002).
\bibitem{Hukushima1996} K. Hukushima and K. Nemoto, J. Phys. Soc. Jpn. {\bf
65}, 1604 (1996).
\bibitem{Metropolis}  N. Metropolis, A. W. Rosenbluth, N. M. Rosenbluth, A. H. Teller, and
E. Teller, J. Chem. Phys. {\bf 21}, 1087 (1953).
\bibitem{DeSimone} C. De Simone, M. Diehl, M. J\"unger, P. Mutzel, G. Reinelt, and G.
Rinaldi, J. Stat. Phys. {\bf 80}, 487 (1995); J. Stat. Phys. {\bf
84}, 1363 (1996).
\bibitem{server} We have used the spin-glass ground-state server of the University of
Cologne where a branch-and-cut algorithm is available online,
http://www.informatik.uni-koeln.de/ls$\_$juenger/index.html.
\bibitem{Roma2008} F. Rom\'a, S. Risau-Gusman, A. J. Ramirez-Pastor, F. Nieto, and E. E. Vogel, in preparation.
\bibitem{Cieplak} M. Cieplak and J. R. Banavar, J. Phys. A. {\bf 23}, 4385 (1990).
\bibitem{PalYou99} M. Palassini and A. P. Young, Phys. Rev. Lett. {\bf 83}, 5126 (1999).
\bibitem{Aspelmeier} T. Aspelmeier, A. J. Bray, and M. A. Moore, Phys. Rev. Lett. {\bf 89}, 197202 (2002).
\bibitem{Palassini2003} M. Palassini, F. Liers, M. J\"unger, and A. P. Young, Phys. Rev. B {\bf 68}, 064413 (2003).
\bibitem{Hartmann4} A. K. Hartmann and A. P. Young, Phys. Rev. B {\bf 66} 094419 (2002).
\bibitem{Hartmannlibro} A. K. Hartmann and H. Rieger, {\em Optimization Algorithms in Physics} (Wiley-VCH, Berlin, 2001).
\bibitem{Blossom} W. J. Cook and A. Rohe, INFORMS J. Comput. {\bf 11}, 138 (1999).
\bibitem{Landry2002} J. W. Landry and S. N. Coppersmith, Phys. Rev. B {\bf 65}, 134404 (2002).
\bibitem{numerical} W. H. Press, S. A. Teukolsky, W. T. Vetterling, and B. P. Flannery, {\em Numerical Recipes in C} (Cambridge University Press, 1992).
\bibitem{Schramm2000} O. Schramm, Isr. J. Math {\bf 118}, 221 (2000).
\bibitem{Flory} P. J. Flory, {\em Principles of Polymer Chemistry} (Cornell University Press, Ithaca NY, 1953).
\bibitem{Schramm2001} O. Schramm, Electron. Commun. Probab. {\bf 6}, 115 (2001).
\bibitem{nota} Of course, one cannot exclude the possibility that there exists an algorithm whose ground state sampling is so much biased that it gives a different value of the slope of the fit.



\end{thebibliography}
\end{document}